\documentclass[12pt]{article}
\usepackage{fullpage,amsmath, amsfonts, amssymb, amsthm, natbib}
\usepackage{graphicx, algorithmic, algorithm, enumerate, hyperref, color}
\usepackage{url} 
\usepackage{upquote}
\usepackage{hyperref}
\hypersetup{breaklinks=true}
\usepackage[table]{xcolor}
\usepackage{pifont}
\newcommand{\cmark}{\ding{51}}%
\newcommand{\xmark}{\ding{55}}%

\newcommand{\student}{{\em Student prompt:}}
\newcommand{\copilot}{{\em Copilot suggestion:}}
\newcommand{\school}{the Marshall School of Business at the University of Southern California}

\newcommand{\red}[1]{{#1}}

\begin{document}

\title{\bf Generative AI for Data Science 101: Coding Without Learning To Code}

\author{Jacob Bien and Gourab Mukherjee\thanks{ The authors gratefully acknowledge Arash Amini for an inspiring conversation about GitHub Copilot that first got us interested in its potential as a teaching tool.}
    \hspace{.2cm}\\
    Data Sciences and Operations, Marshall School of Business\\University of Southern California\\
}
\maketitle

\begin{abstract}

Should one teach coding in a required introductory statistics and data science class for \red{non-major} students?  Many professors advise against it, considering it a distraction from the important and challenging statistical topics that need to be covered.  
By contrast, other professors argue that the ability to interact flexibly with data will inspire students with a lasting love of the subject and a continued commitment to the material beyond the introductory course.
With the release of large language models that write code, we saw an opportunity for a middle ground, which we tried in Fall 2023 in a required introductory data science course in our school's full-time MBA program.  We taught students how to write English prompts to the \red{artificial intelligence} tool {\em GitHub Copilot} that could be turned into {\tt R} code and executed.  In this short article, we report on our experience using this new approach.
\end{abstract}

\section{Introduction}
\label{sec:intro}

A perennial dilemma in teaching a one-semester introductory data science course for non-majors is how much time to devote to teaching students to write computer code.  On the one hand, data science only really comes alive when a student can directly interact with data.  If students are unable to directly apply methods to actual data they care about, they end up as spectators on the sidelines.  They miss out on the valuable experience (and the fun) that comes with directly interacting with data.  \red{Furthermore, when they graduate and face real-world data sets, they are left with no experience applying what they have been taught.}  On the other hand, trying to teach the students to code in a required one-semester data science class might be overwhelming and could distract them from learning the core statistical principles that underlie data science.

In Fall 2023, we co-taught ``Data Science for Business," a one-semester, required class in the full-time MBA program at \school. The course provides a first introduction to the fundamentals of data science and statistics within the context of business applications.  The goal is to introduce students to statistical methods and show how they can be applied to glean useful, business-relevant insights from data.  In previous years, the accepted approach to the dilemma described above has been to avoid coding entirely and instead use a ``point-and-click” software platform.  Such systems allow students to take a carefully cleaned data set and perform a limited set of predefined operations.  We think of this as the ``follow-the-tour-guide" approach to visiting a country.  You don’t speak the language, so you see only what the tour guide has decided it will be feasible for you to see; you see the sights but do not get to chat with the locals.  The other extreme is the ``learn-the-language" approach, which makes for an exciting trip personalized to your interests; however, this admirable route is simply too much effort for most of us.

With the advent of large language models that can produce computer code, we felt we saw a satisfying resolution to this dilemma.  This is the “translator-in-your-ear” approach to travel.  It’s the happy medium, where we can communicate directly with the locals without learning the language.

We seized on this opportunity in Fall 2023 and used the generative \red{artificial intelligence} (AI) tool called GitHub Copilot \citep{GitHubCopilot2022}, which acts like an extremely capable auto-complete system for code.  It was built by OpenAI and GitHub and can be thought of as ChatGPT trained to speak in code rather than English (and fully integrated into a coding environment rather than on a website).   While it is marketed to software developers as an ``AI pair programmer" that can improve their productivity, we used it differently, as an English-to-code translator for students who have never learned to code.  

Here’s an example:

\begin{quote}
\student 

\verb|# Load the data in housing-prices.csv|

\copilot

\verb|housing_prices <- read_csv("housing-prices.csv")|
\end{quote}

Or another:

\begin{quote}
\student

\verb|# Fit a logistic regression model to predict `slow_to_sell`|

\verb|based on `year_built` and `neighborhood`|

\copilot

\verb|housing_prices_model <- glm(slow_to_sell ~ year_built + neighborhood, |

\verb|                            data = housing_prices,| 

\verb|                            family = "binomial")|
\end{quote}

In the latter example, one can observe the great simplification that comes from not having to teach syntax. \red{Understanding generalized linear models and the binomial family's connection to logistic regression is beyond the scope of our course's learning objectives, so in our case this simplification is particularly welcome.}

In this time of great enthusiasm about generative AI, we are certainly not alone in thinking about the possibilities of large language models for teaching data science.  Most of this work has centered on the use of ChatGPT in the classroom \citep[see, e.g.,][and references therein]{ellis2023new}.    One of the most relevant examples to our own is Robert Bray’s Spring 2023 class at Northwestern, also for MBA students, in which he had the students use ChatGPT for generating {\tt R} code \citep{bray2023lessons}.  His paper provides an excellent overview of the experience.  While our approach is similar in spirit to his, we opted for using Copilot rather than ChatGPT because we were concerned that making the class ``open ChatGPT" would open the Pandora’s box too wide---that students would use ChatGPT not just for generating code but for providing a complete solution to a problem.\footnote{\red{Appendix~\ref{app:ai-policy} includes the AI usage policy in our syllabus.}}  GitHub Copilot being embedded in a coding environment makes it less convenient for students to try to get the AI to guide the solution and makes it more likely that students conceive of the AI as a tool for generating code rather than as a tool for reasoning.  Our goal was to use AI to obviate the need for discussing code syntax while leaving all other elements of the class intact.

Of course, pre-AI there is another approach to allow students to use {\tt R} without discussing code. In this approach, the instructor provides code snippets and has students slightly edit the code (e.g., change the name of the data file being loaded, alter the names of the variables, etc.).  Compared to this approach, using GitHub Copilot seems like a much more elegant solution.  First, students using GitHub Copilot are not being asked to scrutinize code (more on this is described in Section~\ref{sec:teaching}).  Second, this copy-and-paste approach is nearly as limited as the ``follow-the-tour-guide'' approach of using {\tt JMP}/{\tt SPSS}/{\tt Minitab}.  It confines a student to only those operations that were encountered in class.

\red{In Table~\ref{tab:compare}, we rate a wide variety of common approaches to software in teaching data science across five characteristics.  The Copilot approach that we describe in this paper is the only method that is favorable across all these characteristics.}

\begin{table}
    \centering
    \scriptsize
 \rowcolors{2}{gray!15}{white}
 \begin{tabular}{rccccc}
    \hline
        	 & \textsc{Work with} & \textsc{Easy to} & \textsc{Extends to methods} & \textsc{Generalizes to} & \textsc{Used in practice} \\ 
        \textsc{} &  \textsc{data?} & \textsc{ teach?} & \textsc{not covered in class?} & \textsc{other languages?} & \textsc{by data scientists} \\ 
        \textsc{No software} & \xmark & \cmark & \xmark& \xmark& \xmark\\ 
        \textsc{Point-and-click} & \cmark & \cmark & \xmark& \xmark& \xmark\\ 
        \textsc{Modify code chunks} & \cmark & \cmark & \xmark& \xmark& \xmark\\ 
        \textsc{Excel} & \cmark & \cmark & \cmark & \xmark& \xmark\\ 
        \textsc{Teach coding} & \cmark & \xmark& \cmark & \xmark& \cmark \\ 
        \textsc{Copilot} & \cmark & \cmark & \cmark & \cmark & \cmark \\ \hline
    \end{tabular}
    \caption{{\em \red{A comparison of six approaches to handling software in data science courses.  Copilot refers to the approach we describe in this paper.  It is the only approach that is favorable across all five characteristics.  In particular, students learning Copilot with R can very easily switch to using Copilot with Python or any other common language.}}}\label{tab:compare}
\end{table}

In Section~\ref{sec:teaching} we discuss the details of our approach.  In Section~\ref{sec:challenges-benefits} we describe the challenges and benefits that became evident through teaching with this approach.  We conclude in Section 4 with some reflections and recommendations.

\section{Teaching to Code Without Teaching Syntax}
\label{sec:teaching}

At the beginning of the semester (August 2023), we had our students sign up for the {GitHub Student Developer Pack} \citep{GitHubStudentPack2023} which gave them free access to GitHub Copilot.  Simultaneously, they installed  {\tt R},  Visual Studio Code, and the {\tt R} and Copilot extensions within Visual Studio Code.  Mid-semester, RStudio introduced GitHub Copilot integration \citep{posit2024}, and we had students use this as well.  The setup process for RStudio proved much simpler for our students, so in the future we would plan on using RStudio.

Most of our students had never coded before, so we began with a brief description of the interactive {\tt R} console presenting it as a fancy calculator and demonstrating little more than arithmetic and the idea of assigning values to variables.  We next introduced {\tt .R} files as a simple way to save and share one's commands so that an analysis can be repeated later in a fresh environment.  From there, we introduced the hash sign along with the idea of comments.  At this point, the focus shifted from the programming language to the idea of prompting Copilot effectively.

We strived to emphasize throughout the class that writing effective Copilot prompts is a skill in itself.  To develop this skill, we found a few key principles useful to convey:

\begin{enumerate}[\em{Principle} 1:]
    \item Be specific.  Do not expect Copilot to read your mind.
    \item Think about the context window.
    \item Break down complex operations into simpler steps.
\end{enumerate}

Suppose we are working with a data set of one-bedroom condo sales listings that contains the listing price and various other variables describing a condo, such as its square footage and number of bathrooms.
Consider a common pitfall in which a student would get frustrated that Copilot was not responding as hoped for based on a prompt such as
\begin{equation}\label{vague-prompt}
\verb|# plot the data.|
\end{equation}

The trouble with this prompt is that it is not at all specific (Principle 1).   A step in the right direction would be

\begin{verbatim}
# plot price versus number of bathrooms for all the condos.
\end{verbatim}

However, a problem with this prompt is that it is expecting Copilot to read our minds (Principle 1) and know the names of the variables.  Copilot will come up with something for the variable names---{\tt price} and {\tt num\_bath}, perhaps---but these may not be correct.  To work effectively with Copilot requires an understanding of its context window (Principle 2): that is, what information is Copilot taking in as a prompt when generating the next line of output?  The details of the context window are actively changing and application specific, including how many tokens\footnote{\red{A ``token'' refers to the basic unit that is generated by a large language model, generally described as longer than a character but sometimes shorter than a word. OpenAI offers an interactive tool that is useful for getting a better sense of how text gets broken into tokens: \url{https://platform.openai.com/tokenizer}.}} are used (an 8000 token context window was announced in the August 28th entry of \citealt{GitHubBlog2023}; \red{for more on the context window, see \citealt{GitHubBlog2023}}) and whether only open code files are used or whether the console output is included as well.  But two important points for our students in Fall 2023 was to realize (i) that Copilot did not have access to the data set itself, even when loaded in the environment, and (ii) the prompt was not just the most recent line being typed but included the other lines in the file.  Thus, the above prompt would be inadequate unless the variable names ``PRICE" and ``BATHS" had been mentioned earlier in the {\tt .R} file.  In such a case, Copilot would be able to map ``price" to ``PRICE" and  ``number of bathrooms" to ``BATHS".  However, even the prompt 

\begin{verbatim}
# plot PRICE versus BATHS
\end{verbatim}

\noindent is leaving too much leeway to Copilot.  In particular, the student is relying on Copilot to decide what type of plot to make.  Copilot might not know whether these variables are numerical or categorical, and thus might not choose the best kind of plot.  The student, not Copilot, should think through what types of variables are being plotted and what type of plot makes sense:

\begin{verbatim}
# make a scatter plot of PRICE versus BATHS
\end{verbatim}
\noindent would be a better prompt, or even

\verb|# make a scatter plot with PRICE on the y-axis and BATHS on the x-axis,|

\verb|  and label the x-axis as "Number of Baths" and the y-axis as "Price (in|
\vspace{-0.4cm}
\begin{equation}\label{most-specific-prompt}
\verb|thousands of dollars)".  Include 0 in the range of the y-axis.|
\end{equation}

In teaching the class, we were careful to avoid using our knowledge of {\tt R} or getting into R-syntax discussions unnecessarily.  Rather, we would demonstrate the refinement of a prompt to convey that writing prompts is a skill and that Principles 1--3 are useful guides.

Principles 1 and 3 are indicative of an important point, which is that one can still teach coding principles without teaching code syntax.  If we ask a student to compute the test-set mean squared error of a regression of price on baths, the student would need to think through the individual sequence of specific steps needed (forming a train-test set split, fitting the model on the training data, making predictions on the test set, calculating the mean squared error using these test set predictions).

Other skills of programming are also still highly relevant---for example, coming up with ways of testing the correctness of the code.  In the context of our class, this would include making a scatter plot, noticing an outlier or data oddity, and then prompting Copilot to try to understand it.  For example, suppose a scatter plot in the housing data reveals one condo that has far more bathrooms than seems plausible.  The student might follow up with the Copilot prompt,

\medskip

\noindent \verb|# print out the row where BATHS is greater than 10|.

\medskip

\noindent The student could then investigate the condo with more than ten bathrooms to decide whether this is a data entry error or whether there is some other explanation for this surprising value.

\section{Challenges and Benefits}
\label{sec:challenges-benefits}

Our choice to use Copilot was an experiment built on the assumption that AI tools will be a permanent part of the landscape for our students after graduation and thus worth teaching.  Going into the semester, we were optimistic but aware that we were in for some surprises.  Here we share some of the challenges and benefits we encountered, which we hope will be useful to others wishing to incorporate these tools into their own teaching.

\subsection{Challenges}

Principles 1--3 described in the previous section were developed mid-semester in response to various challenges we encountered.  One challenge that was particularly disconcerting to students and us alike was grappling with the fact that Copilot’s outputs are random.  Imagine our alarm when we first realized (in front of a classroom of 60+ students) that the prompt that had worked perfectly for us while doing class preparation was no longer generating the same response during class.  In our case, we were teaching three back-to-back sessions, and we found prompts that “worked” earlier in the day would sometimes start generating different and \red{sometimes worse} responses.  Likewise, with 60+ students following along, one could often have 20 students getting something different from the rest.  Principles 1 and 2 are essential for adapting to this challenge.  Returning to the example prompts in the previous section, if one writes the prompt~\eqref{vague-prompt}, there are indeed a large number of reasonable responses that Copilot can give, and so one should not be surprised to find a plethora of outputs produced.  On the other hand, the prompt~\eqref{most-specific-prompt} leaves so little wiggle room that it is unlikely that there will be any variability in the Copilot generated code.

Despite our best efforts to emphasize this important point, some students did not appear to fully grasp the problem with vague prompts, and this would lead to frustration during quizzes (``What I did was working before the quiz!  Why did Copilot stop working?!").  The fact that Copilot when given a vague prompt would {\em sometimes} do the right thing tended to undermine our repeated messaging to write specific prompts and not expect Copilot to read our minds.

An amusing example of vague prompts sometimes leading to the intended behavior emerged when working with classic data sets.  For example, we used the well known ``wine" data set from the UCI data bank \citep{misc_wine_quality_186}.  After writing the prompt ``{\tt\#~load the file named wine.csv}", Copilot would go on to auto-suggest that we predict the wine’s quality based on the other variables.  This was exactly what we intended to be doing.  To students, such behavior would look like Copilot was able to see the names of variables in the data set (and seem to have an opinion about interesting modeling tasks).  In fact what was happening is that the wine data set had been used so extensively across GitHub (the training set for Copilot) that it was simply recapitulating the huge number of analyses of this particular data set that it had seen.  Sometimes such behavior would lead to confusion.   For example, when working with some housing price data, Copilot might assume, incorrectly, that there was a variable such as {\tt square\_feet} (rather than {\tt SQFT}) before any mention of this variable was made.  \red{For more on this example, we encourage the reader to turn to Appendix~\ref{app:full-session}, in which we provide a full walk-through of a Copilot session, showing screenshots interspersed with commentary.}

\red{One notable form of inconsistency in code generated by Copilot is whether it opts for {\tt tidyverse} \citep{tidyverse} or base {\tt R} in a given analysis. (This reflects the inconsistent use of the two in the code that Copilot was trained on.) To prevent this form of variability (and to reflect our own coding preferences), we would regularly have students start {\tt .R} files with the Copilot command ``{\tt\# load the tidyverse {\tt R} package.}"  This one line in the file would generally lead Copilot to opt for {\tt readr::read\_csv()} instead of {\tt read.csv()} and to use {\tt dplyr} and {\tt ggplot2} without the students needing to know anything about those packages.}

Another challenge in using Copilot was the lack of transparency about its inner workings and the fact that these inner workings could be changing over time.  For example, Principle 2 requires understanding the details of the context window used by Copilot; however, to our knowledge there is no definitive documentation that spells out these details unambiguously, which left us sifting through blog posts, change logs, and GitHub staff responses to users on community forums. \red{Furthermore, another downside of Copilot is that it does not allow for offline coding.}

\subsection{Benefits}

We were impressed by a number of benefits that emerged through our teaching Copilot to our students.  The primary benefit was that it put students in the driver’s seat, allowing them to interact with data guided by their curiosity rather than constrained by a limited set of operations they had been taught.  The absence of a prespecified list of operations highlighted for the students the boundless nature of the tool.

A story from the semester conveys the value of this benefit.  In one class, we were fitting different regression models to predict the quality of a wine based on its physicochemical properties.  The response is the median score of three experts and is an integer ranging from 0 to 10 \citep{cortez2009modeling}. The purpose of the class was to demonstrate how to compute \red{test-set} mean-squared errors to assess how different regression methods compared in terms of predictive performance.  A scatter plot of the actual versus predicted response revealed a horizontal stripe pattern.  We discussed as a class how this was due to the observed response being integer-valued (while the predicted values were not similarly constrained).  Our conclusion from this exercise was that stepwise regression did not appear to have made much of a difference in prediction error relative to a regression model with all predictor variables included.  After class, a student came up to one of us with his laptop.  He showed that if one rounds the predictions to the nearest integer, one gets a substantial reduction in test error.  This was a thrilling moment for us in revealing the power of the Copilot approach.  We had never mentioned that there was a function called “round” in  {\tt R}.  The student had gone from a hunch (``I wonder if rounding would help?'') to an implementation with complete ease.  While one might expect such a modification to be straightforward for a student who has been learning  {\tt R},  \red{this was striking to us because} this student had never used {\tt R} before---and our class was certainly not teaching the {\tt R} language either.  The student in fact had never programmed before and had entered the semester unconvinced that the required data science course would be of any interest or relevance to him. \red{To us, the key insight in this story is that Copilot lowers the barrier to experimentation: if you can precisely express what you want to do in English, you can probably produce code to do it fairly easily (without the levels of motivation required to do web searches for {\tt R} commands).}

\red{As the example above demonstrates, we found that our use of Copilot led students to take an expansive view of what they could accomplish in code (rather than being confined to a small set of operations).}  Several of our students told us they had started using Copilot in working with their own business data for data manipulation and analysis tasks.

\red{Another benefit of the Copilot approach is that it is based in code (in contrast to typical point-and-click or Excel-based approaches), allowing instructors to impart reproducible data analysis practices.  For example, we taught our students to use Copilot to create {\tt R Markdown} files and discussed the importance of setting the seed before doing analyses that use (pseudo)randomness such as when one splits a sample into training and test sets.  From the perspective of reproducibility, it is interesting to note that while the {\em final product} of using Copilot is a {\tt.R} or {\tt.Rmd} file of code,  which records a data analysis in a reproducible way, the {\em process of prompting} Copilot to generating code is not reproducible.  This is analogous to how the same human on two different days might solve the same problem in two different ways, a factor that is not relevant when considering the reproducibility of an analysis.}

Finally, a striking benefit of the Copilot approach is that it is language agnostic.  While we had our students using {\tt R} throughout the semester, the prompting skills we were teaching would be unchanged if we had instead done this in Python.

\section{Reflections and Recommendations}
\label{sec:recommendations}

We found this Copilot approach to teaching introductory data science to be quite successful for our students and the class we were teaching. That said, we do not necessarily recommend this approach in other contexts.  For example, if this had been a data science elective course consisting of students with the background, time, and interest to learn to code, we would be reluctant to use Copilot.  We imagine that using Copilot in such a class could undermine a student's learning process (analogous to giving young students calculators before they have learned how to do arithmetic).

In describing our approach to colleagues, a common objection was the following: If students do not know how to read code, then how could they possibly know if the code produced by Copilot is correct?  This objection is usually accompanied by stories of the person’s experience using ChatGPT to generate code.  In our opinion, this concern can be alleviated by teaching students the practice of checking the reasonableness of outputs.  For example, after calculating a new variable, one can prompt Copilot to print the five largest and five smallest values of this new variable.  Students can be encouraged to make plots of the data and other outputs after each step to verify that the calculations are matching the intention.  This is an imperfect process that is unlikely to be sufficient for complex programming projects; however, for an introductory data science course such as ours, this emphasis on making many plots and scrutinizing them critically is completely consistent with standard data analysis principles that we would be teaching anyway (such as not taking the correctness of one’s data set for granted).

When teaching this way in the future, there are a few changes we may consider making.  First of all, since our students have never coded before, they seem to be a bit hazy on the distinction between {\tt R} and GitHub Copilot (and RStudio).  For example, students would say, “Copilot generated this plot.” It felt overly pedantic to correct our students to say, “Copilot generated this {\tt R} code that we ran to generate this plot.”  We might consider teaching the first class in {\tt R} without Copilot in order to cement the distinction.  Another consideration is when to introduce {\tt R} Markdown.  {\tt R Markdown}  is particularly valuable for our MBA students since they will be in roles in which clear communication is essential.  We introduced {\tt R Markdown}  toward the end of the semester after many weeks of writing simple {\tt R} files.  Copilot works seamlessly with {\tt R Markdown} .  Here’s an example:

\begin{quote}
\student

\verb|We next make a plot |

\copilot

\verb|```{r}|

and then after this suggested output is accepted by the student (by pressing the tab key), Copilot would suggest finishing the code chunk with

\begin{verbatim}
ggplot(data, aes(x = x, y = y)) +
  geom_point()
```
\end{verbatim}
\end{quote}

However, the transition in the class from {\tt .R} files to {\tt R Markdown}  was a bit awkward since the meaning of the hash sign ``{\tt\#}" changes.  The students had become accustomed to every Copilot prompt starting with ``{\tt\#}" (because their prompts were always comments in {\tt .R} files).  However, in {\tt R Markdown} , the ``{\tt\#}" designates section headers, and one writes prompts in plain text without any prefix (as in the example above).  It would be tempting to teach {\tt R Markdown}  from the start and skip {\tt .R} files entirely.  When teaching {\tt R} coding without Copilot, starting with {\tt .R} files before {\tt R Markdown}  files makes sense to us; however, since our approach to using Copilot is to have the student write a prompt in English and have Copilot respond with some {\tt R} code, this approach aligns naturally with the format of writing an {\tt R Markdown}  file.  Therefore, we might teach {\tt R Markdown}  from the beginning in the future.

\bibliographystyle{agsm}
\bibliography{refs.bib}

\appendix
\red{\section{Course Information}}
\red{\subsection{Background and Course Details}}

``GSBA-524: Data Science for Business'' is a compulsory class for all first-semester, full-time MBA students. It has three sections usually taught by the same set of instructors with around 65 students in each. Each section meets on average 3 hours per week.  The prerequesite is ``a basic understanding of descriptive statistics,'' and a summertime several-hour ``jumpstart'' is offered to any students who may not have this background.

In Fall 2023, GM taught from August 21--October 10, covering descriptive statistics, data visualization, introductory probability, sampling distributions, the central limit theorem, confidence intervals, hypothesis testing, and simple and multiple regression; JB taught from October 16--December 13, covering interaction models, inference in regression, variable selection, predictive analytics, neural networks, logistic regression, classification evaluation, and unsupervised learning.   These topics are presented to students within the context of business applications.  In Fall 2023, we left the course topics and structure unchanged, but switched from using Radiant \citep{radiant}, which is a browser-based point-and-click app based on {\tt shiny} \citep{shiny} and the {\tt R} programming language \citep{R}, to the GitHub Copilot approach described in the main paper.

There were 31 class sessions across the semester (each being 1.25 hours long), six homework assignments, six quizzes, a midterm, and a final exam (along with some additional ``participation assignments'').  By the end of the course, students were expected to turn in an {\tt R Markdown}  file (and the rendered {\tt .html} output) for homework assignments.  For quizzes and exams, students were expected to use GitHub Copilot (e.g. a typical quiz question might ask a student to download a {\tt .csv} file, read in the data, make a plot, and answer a question based on the plot).  For the final exam, the first component, which did not involve computation, was taken on paper with no technology allowed.

In addition to holding a weekly office hour (as is typical for this course), we added two hours of PhD office hours each week, so students could get additional assistance with installing and using GitHub Copilot.

\red{\subsection{AI Policy Communicated to Students}\label{app:ai-policy}}

We included the following AI usage policy in our course syllabus (which is a modification of a template provided by our college):

\begin{quote}
``In this course, you will be using GitHub Copilot to create code for data analysis in class, for assignments, on
quizzes, and on exams. However, no other generative AI tools are permitted on quizzes and exams. The use
of non-Copilot AI tools on assignments, while not strictly forbidden, is discouraged since the purpose of
assignments is for you to learn the material and using these tools might negatively affect your learning process.''
\end{quote}

\red{\subsection{Setup Instructions}}
\label{app:setup}

While the installation details may of course change over time, there are two basic components to the setup of GitHub Copilot with {\tt RStudio}:

\begin{enumerate}
    \item {\em Getting GitHub Copilot access.}  This was free for students and instructors in Fall 2023 (and is still free as of Fall 2024) through the ``GitHub Education'' program.  This involves verifying an academic email address and showing proof such as a university identification card.
    \item {\em Downloading and installing {\tt RStudio}.}  Within {\tt RStudio}, one goes to Tools $>$ Global Options and then finds the tab labeled ``Copilot.''  For more instructions, see \url{https://docs.posit.co/ide/user/ide/guide/tools/copilot.html}.
\end{enumerate}

\red{\section{A Session with Copilot Demonstrating Principles 1 and 2}\label{app:full-session}}

In this appendix, we provide a full walk-through of a Copilot session that demonstrates the importance of Principles 1 and 2.  We intersperse screenshots (taken from our screen recording available at {\footnotesize\url{https://faculty.marshall.usc.edu/jacob-bien/copilot/specific-prompts.mov}}) with commentary.

\medskip
\noindent {\bf Background:} So far in this session, we have created a file {\tt redfin.R} that has code, which we have already run in the console, that (i) loads the {\tt tidyverse} {\tt R} package, (ii) reads in {\tt redfin.csv} resulting in a data frame named {\tt redfin}, and (iii) prints the names of the variables in this data frame.  We wish to make a scatter plot that shows the relationship between the price of a condo and its square footage (the file {\tt redfin.csv} was created with students in class by downloading condos that were currently for sale in Los Angeles from the website \url{https://www.redfin.com/}).

\medskip
\noindent {\bf First attempt:} As described in the paper, our approach is to write comments that will prompt Copilot to generate code.  In our first screenshot, we begin typing the instruction ``\verb+# Make a scatter plot of+".  We can see in gray that Copilot suggests completing this prompt with ``\verb+price vs. square footage+":

\begin{center}
\includegraphics[width=0.7\linewidth]{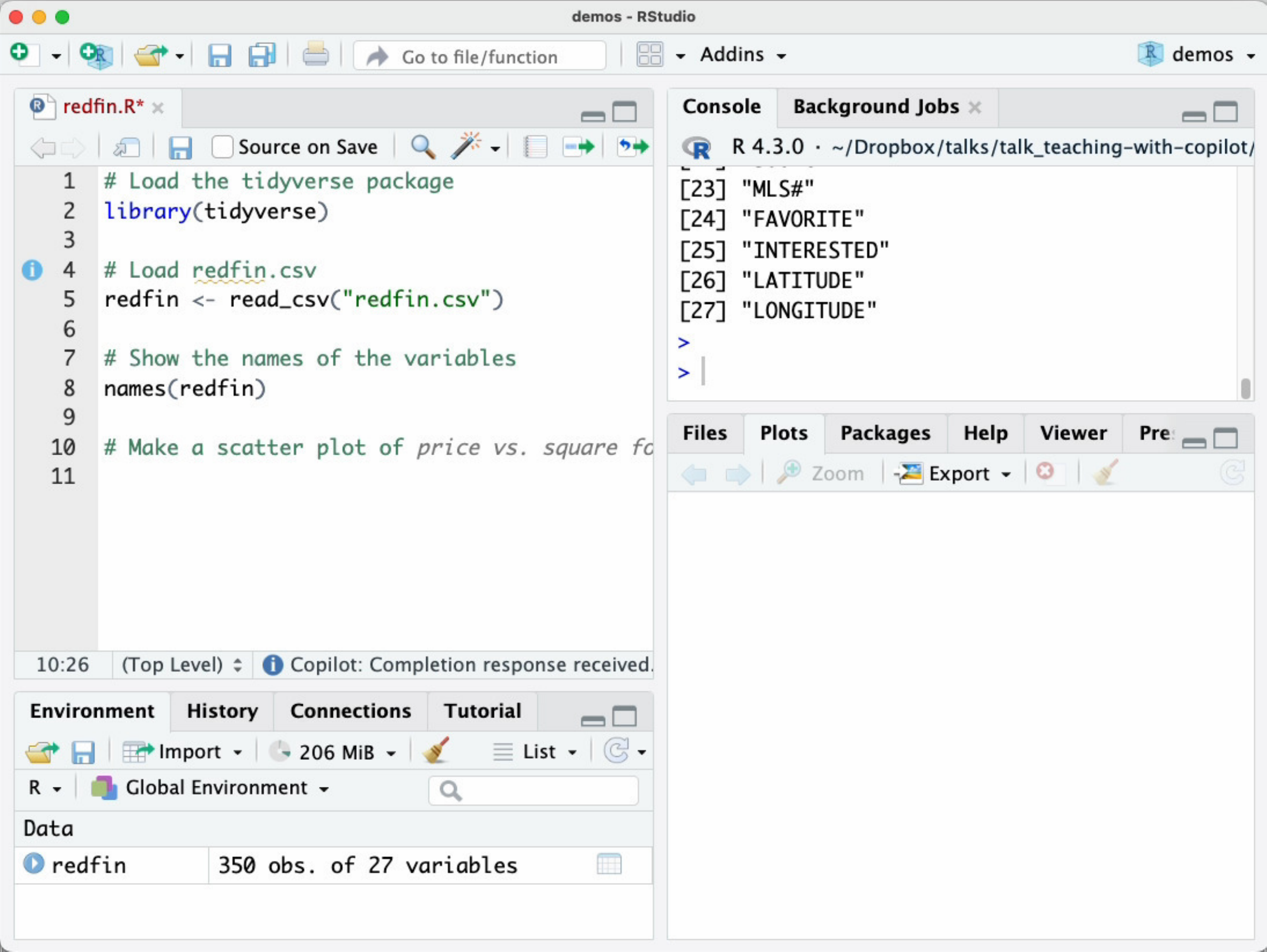}
\end{center}

As it happens, our goal was to make a scatter plot with a condo's price on the y-axis and its square footage on the x-axis, so it feels like Copilot is reading our mind!  What could explain this amazing behavior?  Copilot was trained on a vast quantity of code files, many of which included data analyses.  Previous analyses that included the word {\tt redfin} tended to be real estate data analyses, which often involve predicting price based on other variables (and square footage is certainly a common factor to consider in such contexts). We naively accept the gray prompt by pressing the [TAB] key.  This in turn leads Copilot to suggest {\tt ggplot2} code in the next two lines, which we also accept with presses of the [TAB] key.  Its choice of {\tt ggplot2} code rather than base {\tt R} plotting functions is influenced by line 2, where {\tt tidyverse} had been loaded.  We next run these lines of generated code in the console, which generates the error shown in the next screenshot: 

\begin{center}
\includegraphics[width=0.7\linewidth]{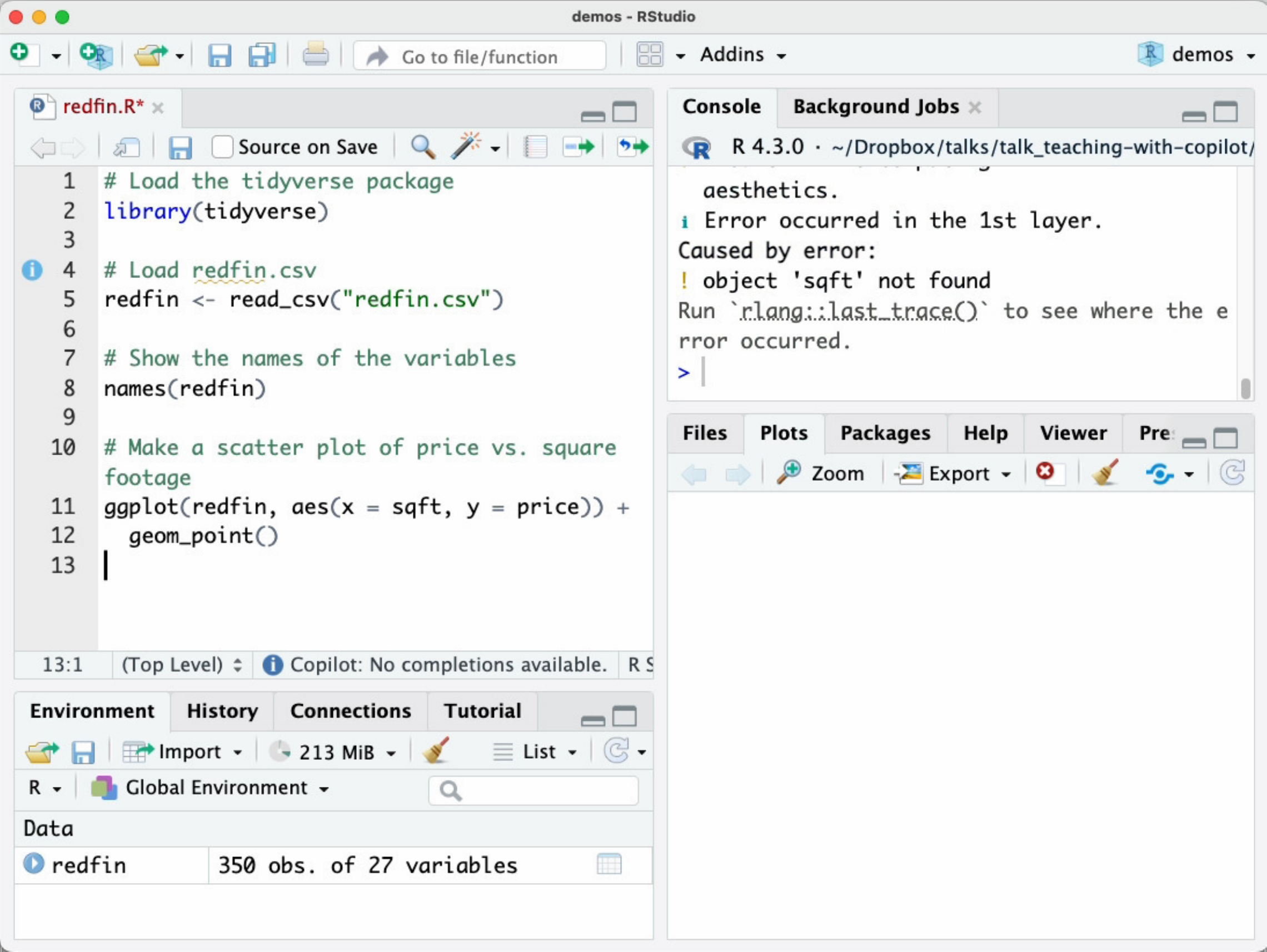}
\end{center}

The error message is ``object `sqft' not found".  Looking at line 11 in the code file, we see that Copilot has made the incorrect assumption that the name of the square footage variable is ``\verb+sqft+''.  What went wrong here?  Recall that Copilot makes its suggestions based on the information available to it, which is known as the ``context window.''  The context window when it generated lines 11 and 12 would have been lines 1--10 of {\tt redfin.R}.  Notice that the correct variable names ({\tt SQUARE FEET} and {\tt PRICE}) do not appear in this file.  Interestingly, RStudio and {\tt R} both have access to the variable names since the data frame is in the global environment and we have even previously printed the variable names in the console when we ran line 8 earlier.  The next screenshot shows the correct variable names that were earlier outputted from the console:

\begin{center}
\includegraphics[width=0.7\linewidth]{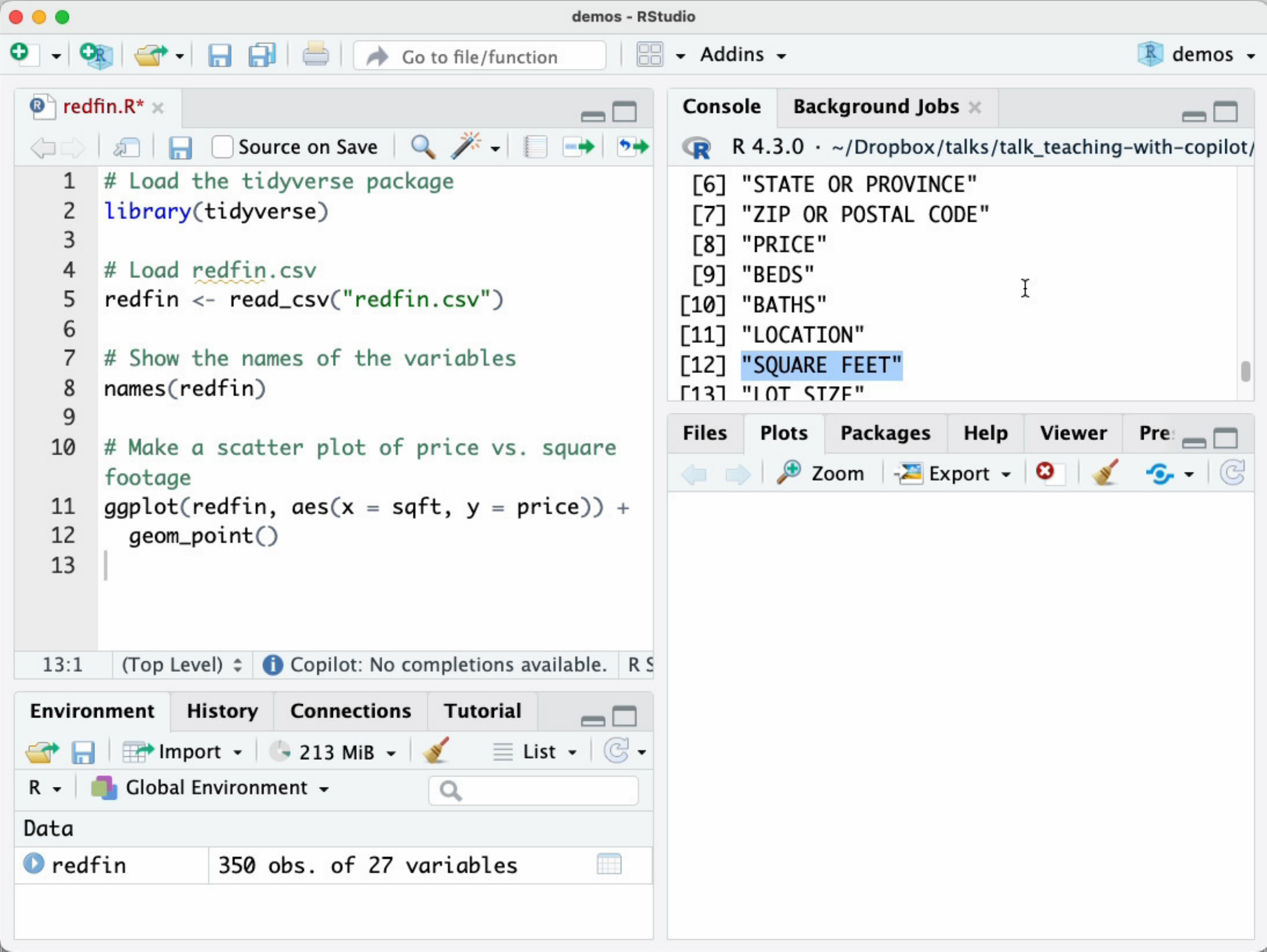}
\end{center}

However, Copilot does not appear to have access to either the global environment or to the output from the console.  In other words, the global environment and the output from the console are not part of the context window.\footnote{This choice could of course change in future iterations of the software.}

\noindent{\bf Second attempt:} Armed with this new knowledge, we rewrite our prompt in line 10.  We follow Principle 1 and are {\em specific} in writing the exact names of the variables rather than leaving this to be guessed by Copilot:

\begin{center}
\includegraphics[width=0.7\linewidth]{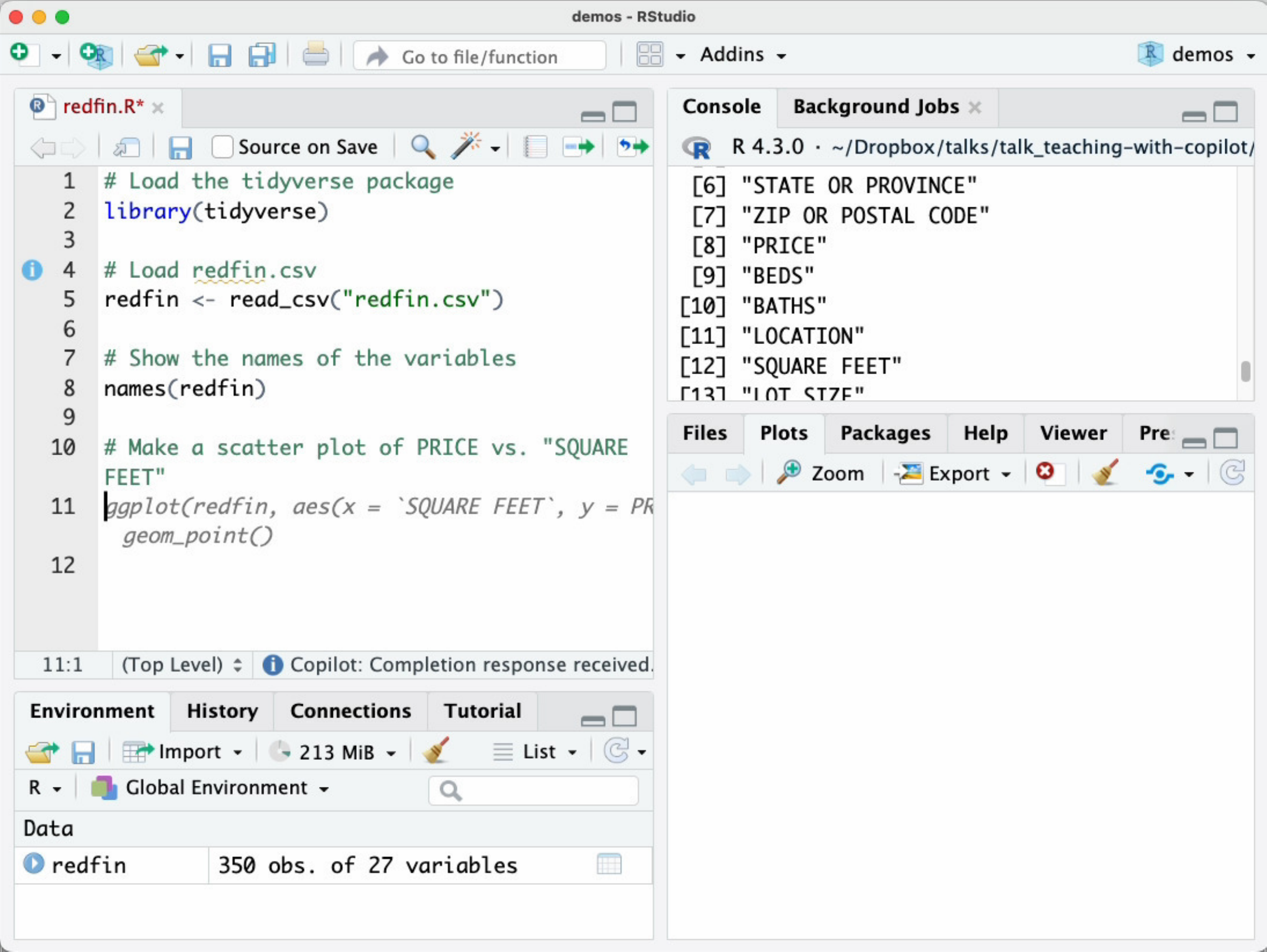}
\end{center}

When we run this code, we get the desired plot:

\begin{center}
\includegraphics[width=0.7\linewidth]{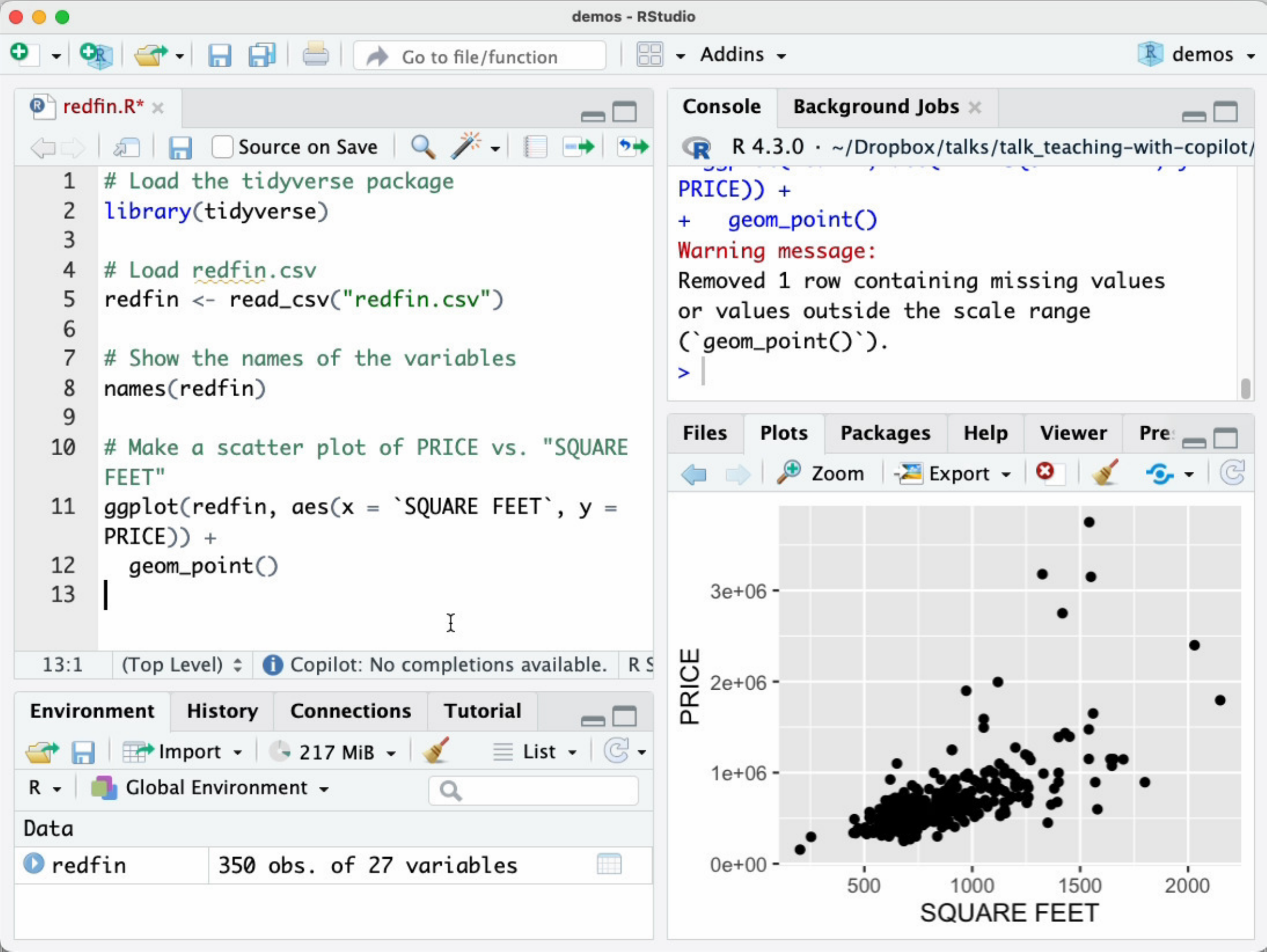}
\end{center}

\noindent{\bf Investigating the plot:}
Having successfully generated this plot, we notice a condo that is priced above \$3.5 million.  This seems quite expensive, so we would like to make sure that this is not a data error.  We form a prompt (in line 14) asking to find that condo in the data set, and Copilot yields in line 15 a correct code suggestion:

\begin{center}
\includegraphics[width=0.7\linewidth]{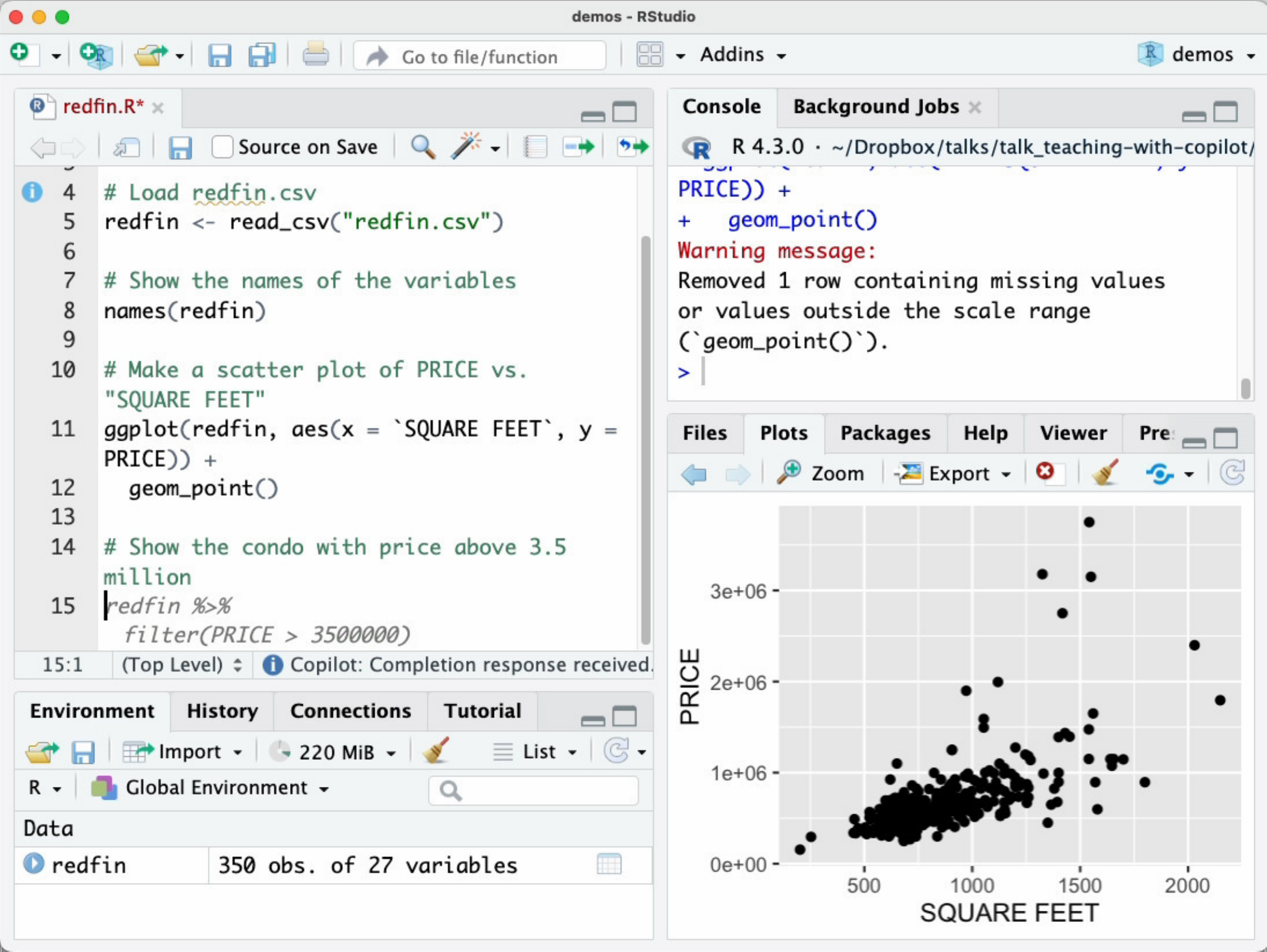}
\end{center}

It is instructive to note that unlike earlier, we can now get away with referring to ``\verb+price+" rather than needing to write ``\verb+PRICE+''.  What is the difference from earlier?  The context window has changed (Principle 2).  In particular, when line 15 is being generated, Copilot has access to lines 1--14, which include a reference to ``\verb+PRICE+'' in lines 10 and 11.    To get a more useful output, we ask for the condo's address:

\begin{center}
\includegraphics[width=0.7\linewidth]{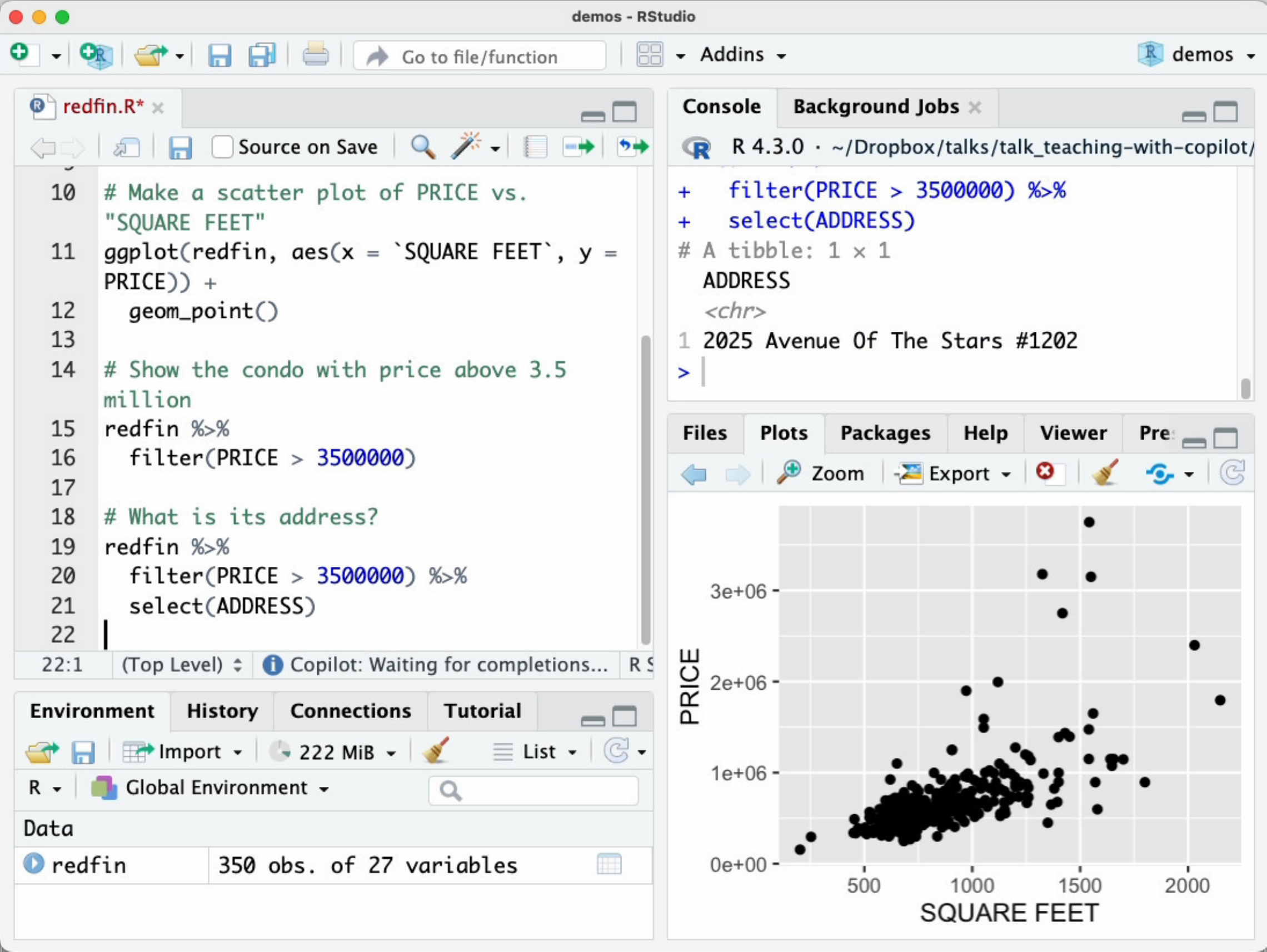}
\end{center}

This yields a specific address that we can Google (leading us to determine that this is not a data error, just a very expensive condo!).  In this case, Copilot correctly guessed that the name of the variable was ``\verb+ADDRESS+".  From line 10, it appears to have inferred that all the variable names are in all-caps, regardless of how they are referred to in the prompts.  Another correct inference it made was that our prompt's use of ``its" was referring to the condo that had been returned in lines 15--16.

\bibliography{refs.bib}
\end{document}